\documentclass[aps,prl,reprint,groupedaddress]{revtex4-1} 
\usepackage{graphicx}
\usepackage{amsmath, amsxtra, amssymb, mathtools,amssymb}
\usepackage{isomath, nccmath, xspace, siunitx}
\usepackage{textcomp}
\usepackage{dsfont} 
\usepackage[english]{babel}

\newcommand{\ket}[1]{\ensuremath{\lvert #1 \rangle}\xspace}%
\newcommand{\bra}[1]{\ensuremath{\langle #1 \rvert}\xspace}%
\newcommand{\abs}[1]{\ensuremath{\lvert #1 \rvert}\xspace}%

\newcommand{\Rb}{\ensuremath{^{87}\text{Rb}}\xspace}%
\newcommand{\Rbk}{\ensuremath{R_{\text b}}\xspace}%
\newcommand{\alat}{\ensuremath{a_{\text{lat}}}\xspace}%

\newcommand{\gtwo}{\ensuremath{g^{(2)}}\xspace}%
\renewcommand{\vec}{\vectorsym}

\begin{document}

\title{Finite-range interacting Ising quantum magnets with Rydberg atoms in optical lattices - From Rydberg superatoms to crystallization} 

\author{Peter Schauss}
\affiliation{Department of Physics, Princeton University, Princeton, New Jersey 08544, USA}

\date{\today}

\begin{abstract}
Finite-range interacting spin models are the simplest models to study the effect of beyond nearest-neighbour interactions and access new effects caused by the range of the interactions. Recent experiments have reached the regime of dominant interactions in Ising quantum magnets via optical coupling of trapped neutral atoms to Rydberg states. This approach allows for the tunability of all relevant terms in an Ising Hamiltonian with $1/r^6$ interactions in a transverse and longitudinal field. This review summarizes the recent progress of these implementations in Rydberg lattices with site-resolved detection.
The strong correlations in this quantum Ising model have been observed in several experiments up to the point of crystallization. In systems with a diameter small compared to the Rydberg blockade radius, the number of excitations is maximally one in the so-called superatom regime.
\end{abstract}


\maketitle


\section{Introduction}

Spin Hamiltonians are often introduced as tractable simplifications of realistic condensed matter Hamiltonians \cite{Auerbach1994,Schollwoeck2004,ParkinsonFarnell2010}. Quantum simulation of these spin Hamiltonians can be performed by ultracold atoms in optical lattices \cite{BlochDalibardZwerger2008,Lewenstein2012}. Heisenberg-type spin Hamiltonians have been implemented in optical lattices by second order tunnelling, the so-called superexchange process \cite{Foelling2007,Trotzky2008}. In fermionic systems superexchange  leads to antiferromagnetic correlations \cite{Greif2013,Hart2015}. Progress in site-resolved imaging of ultracold bosonic atoms in optical lattices has enabled the study of spin Hamiltonians with single spin resolution in a very controlled setting \cite{Fukuhara2013a,Fukuhara2013b,Preiss2015} and has recently been extended to fermions allowing to study antiferromagnetic correlations in the repulsive Hubbard model \cite{Parsons2016,Boll2016,Cheuk2016,Brown2016}. The main advantage of these single-site resolution techniques is their single-particle sensitivity in density-snapshots of the full many-body wavefunction and therefore the direct access to spatial correlation functions \cite{Endres2013}. Typically the interactions in these systems are limited to nearest neighbour interactions. Spin systems with power-law interactions are predicted to show new physics not observable in systems with at most nearest neighbour interactions \cite{Baranov2008,Hazzard2014}. Examples include spin glasses \cite{Olmos2012b,Lechner2013,Angelone2016}, quantum crystals \cite{Weimer2008,Pupillo2010} and modified light cone dynamics \cite{Schachenmayer2013,Hauke2013}.

In order to investigate longer range interacting spin models new experimental techniques are required. One direction is to employ dipolar molecules in optical lattices \cite{Yan2013} or magnetic atoms \cite{Paz2013,Baier2016}. Long range interactions have also been implemented with ions in one-dimensional systems \cite{Richerme2014,Jurcevic2014} and extended to two dimensions \cite{Britton2012,Bohnet2016}. Another route is to exploit the power-law interactions between Rydberg atoms \cite{Saffman2010,Comparat2010,Browaeys2016}. While molecules and magnetic atoms are expected to have a very long lifetime in a suitably designed trapping configuration, Rydberg atoms are relatively short-lived but have much stronger interactions. Currently the lifetime to interaction timescale ratio ends up in a comparable range and none of these systems has shown to be superior yet \cite{Hazzard2014}. However, long-range interacting systems based on trapped ions currently show the longest coherence times where analog simulation is possible \cite{Porras2004}.

This review focuses on the approach to use Rydberg atoms to implement quantum Ising systems with beyond nearest neighbour interactions in lattice configurations.
Rydberg atoms are atoms with at least one electron in a highly excited state and exhibit surprisingly long lifetimes of typically few ten microseconds to few milliseconds due to the low overlap of the excited electronic state with the ground state \cite{Gallagher1994}. Their large electronic wavefunction leads to strong  induced dipole-dipole interactions. These van der Waals interactions typically show a $1/r^6$ dependence with distance $r$ in the absence of special resonances. Rydberg states exist for all atoms and therefore constitute a general concept to generate long-distance interactions that is applicable for all atoms.

 This review is organized as follows. It starts with a short introduction to the mapping of the naturally arising Rydberg Hamiltonian to a spin Hamiltonian and then discusses the experimental implementation.
 Thereafter, we discuss systems in the regime of strong interactions, where the interaction range exceeds the system size. As a consequence, the system can only host a single excitation and is well described by a so-called ``superatom". The following part focuses on larger systems where correlations and crystallization have been observed. In the last part future directions are discussed.
 It is not intended to give a general introduction to Rydberg physics here and the reader is referred to other reviews on this topic \cite{Gallagher1994,Saffman2010,Comparat2010,Low2012,Browaeys2016}.

\section{From the Rydberg Hamiltonian to the Ising model}

\begin{figure}
\includegraphics{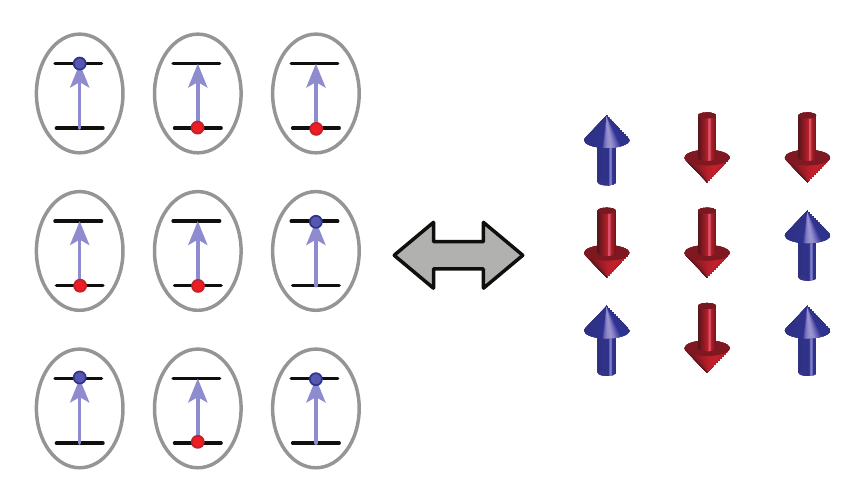}
\caption{Mapping the Rydberg system to a spin Hamiltonian. On the left an array of atoms in the ground state (lower level) is coupled with Rabi frequency $\Omega$  (blue arrows) to the Rydberg state $\ket{e}$ (upper level). On the right this corresponds to the same number of spins, where $\ket{\uparrow}$ corresponds to atoms in the Rydberg state (blue circles) and $\ket{\downarrow}$ to atoms in the ground state (red circles).  \label{fig:mapping}}
\end{figure}

Here we consider a setup where many ground state atoms are coupled with a single laser field to a Rydberg state (Fig.~\ref{fig:mapping}). When coupling atoms with a laser near-resonantly with Rabi frequency $\Omega$, the experimental timescales are maximally on the order of the Rydberg atom lifetime $\tau$. This leads for ultracold atoms to the observation that the motion of the atoms is in most cases negligible for the excitation dynamics. The corresponding theoretical model describing this system is the so-called ``frozen Rydberg gas", in which only the internal electronic degrees of freedom are considered and all atomic motion is neglected~\cite{Anderson1998,Mourachko1998}.
A very important concept for the understanding of these systems is the so-called dipole blockade \cite{Jaksch2000,Lukin2001}. The interactions between Rydberg atoms lead to a relevant resonance shift of the Rydberg excitation line in ground state atoms close to a Rydberg atom. This interaction-induced detuning of the transition frequency suppresses the excitation of Rydberg atoms at small distances and equating the interaction potential, $V(r)$, with the excitation bandwidth gives an estimate for the length scale of this effect. For resonant excitation the bandwidth is typically given by the Rabi frequency and the blockade radius $R_b$ is therefore defined by $V(R_b) = \hbar \Omega$. The blockade constraint in the relative distance between Rydberg excitations by itself leads to strongly correlated many-body states \cite{Robicheaux2005}.

In this review, we focus on systems where single atoms, arranged on a square lattice with positions $\vec i$, in a ground state $\ket{g_{\vec i}}$ are coupled to a Rydberg state $\ket{e_{\vec i}}$. We assume homogeneous Rabi frequency $\Omega(t)$ and detuning $\Delta(t)=\omega_l(t)-\omega_0$ of the coupling for all atoms, both considered time-dependent. The lattice setting  comes with the inherent advantage that molecular loss between atoms can be suppressed by choosing a Rydberg state that has no molecular lines in the range of $\Delta(t)$ of interest and beyond the minimum spacing between two atoms, the lattice constant $\alat$ \cite{Derevianko2015}.

For the basic model no detailed knowledge about Rydberg atoms is required, as the main property of interest here is the interaction between two Rydberg atoms in the same state which can be approximated by a van-der-Waals potential of type $V(r) = -C_6/r^6$ for distances $r \ge \alat$ and large compared to the electronic wavefunction. The interaction potential can become more complicated in the presence of F\"orster resonances which are not discussed here \cite{Forster1946}. More details on Rydberg atoms can be found in the reviews \cite{Saffman2010,Comparat2010,Low2012,Browaeys2016}.
We note that the interactions between Rydberg and ground state atoms are not relevant here under the assumption that the Rydberg state is chosen as not to overlap with ground state atoms on neighbouring sites. Ground state atoms would interact only if two of them were on the same site.

The Hamiltonian describing the system is given in rotating wave approximation by

\begin{equation}
    \label{eq:H}
    \hat{H} = \frac{\hbar\Omega(t)}{2} \sum_{\vec i} \left( \hat\sigma^{(\vec i)}_{eg} + \hat\sigma^{(\vec i)}_{ge} \right)
    -  \hbar \Delta(t) \sum_{\vec i} \hat\sigma^{(\vec i)}_{ee} + \sum_{\vec i \neq \vec j} \frac{V_{\vec i \vec j}}{2} \hat\sigma^{(\vec i)}_{ee} \hat\sigma^{(\vec j)}_{ee} \; .
  \end{equation}
  Here, the vectors $\vec i = (i_{x}, i_{y})$ label the lattice sites in the plane. The first term in this Hamiltonian describes the coherent coupling of the ground and excited states with
     Rabi frequency $\Omega(t)$, with $\hat \sigma^{(\vec i)}_{ge} = \ket{e_{\vec i}}\bra{g_{\vec
        i}}$ and $\hat \sigma^{(\vec i)}_{eg} = \ket{g_{\vec i}}\bra{e_{\vec i}}$.
		The second term takes into account the effect of the laser detuning $\Delta(t)$. The projection operator $\hat \sigma^{(\vec i)}_{ee} = \ket{e_{\vec i}}\bra{e_{\vec i}}$ measures the population of the Rydberg state at site $\vec i$.
		The third term is the interaction potential between two atoms in the Rydberg state $V_{\vec i\vec j} = -C_{6}/r_{\vec i\vec j}^{6}$, with van der Waals coefficient $C_{6}$ and $r_{\vec i\vec j} = \alat |\vec i - \vec j|$ the distance between the two atoms at sites $\vec i$ and $\vec j$.

 By identifying $\ket{g_{\vec i}} = \ket{\downarrow_{\vec i}}$ and $\ket{e_{\vec i}} = \ket{\uparrow_{\vec i}}$ the Hamiltonian can be rewritten as a spin Hamiltonian for which we introduce spin-$1/2$ operators on each site as follows (Fig.~\ref{fig:mapping}): We define $\hat S^{(\vec i)}_{x} = (\ket{\uparrow_{\vec i}}\bra{\downarrow_{\vec i}} + \ket{\downarrow_{\vec i}}\bra{\uparrow_{\vec i}})/2$ and $\hat S^{(\vec i)}_{z} = (\ket{\uparrow_{\vec i}} \bra{\uparrow_{\vec i}} - \ket{\downarrow_{\vec i}} \bra{\downarrow_{\vec i}})/2$ and note that $\hat\sigma^{(\vec i)}_{ee} = \frac{\mathds{1}}{2}+\hat S^{(\vec i)}_{z}$ with $\mathds{1}$ being the identity. The operators $\ket{\uparrow}\bra{\downarrow}$ and $\ket{\downarrow}\bra{\uparrow}$
describe a spin flip from the ground state $\ket{\downarrow}$ to the Rydberg state
$\ket{\uparrow}$ and vice versa, while the operators $\ket{\uparrow} \bra{\uparrow} = \hat
n_\uparrow$ and $\ket{\downarrow} \bra{\downarrow} = \hat n_\downarrow$ represent the local
Rydberg and ground state population, respectively. Using these expressions and neglecting a constant offset, the Hamiltonian becomes:

 \begin{equation} 
  \hat{H} = \hbar\Omega(t)
  \sum_{\vec i} \hat S^{(\vec i)}_{x} + \sum_{\vec i} [\mathcal{I}_{\vec i} - \hbar
  \Delta(t)] \hat S^{(\vec i)}_{z} + \sum_{\vec i \neq \vec
  j} \frac{V_{\vec i \vec j}}{2} \hat S^{(\vec i)}_{z} \hat S^{(\vec
  j)}_{z}\; . 
\end{equation} 
The first two terms of this spin Hamiltonian describe a transverse and longitudinal magnetic field. The former is controlled by the coherent coupling between the ground and the Rydberg state with the time-dependent Rabi frequency $\Omega(t)$. The detuning $\Delta(t)$ determines the longitudinal field and can be used to counteract the energy offset $\mathcal{I}_i = \sum_{\vec j,(\vec i \neq \vec j)} \frac{V_{\vec i \vec j}}{2}$. In finite systems, $\mathcal{I}_i$ has a spatial dependence with negligible consequences besides favouring the pinning of Rydberg excitations at the edge of the system. In infinite systems, $\mathcal{I}_i$ just leads to a constant offset of the detuning.
This spin Hamiltonian has a dynamic aspect in the sense that it requires the presence of driving to a Rydberg state which inherently limits its study to timescales shorter than the Rydberg lifetime. But the interactions can be very strong, and therefore the time evolution fast enough that the properties of the system can be explored before the decay processes become relevant. Experimentally the strength of the interactions can easily exceed the linewidth of the Rydberg excitation line and the maximum achievable Rabi frequency. Therefore the relevant figure of merit for the observation of coherent dynamics is the product of system lifetime $\tau$ and $\Omega$. Values of $\Omega\tau > 100$ are experimentally feasible \cite{Jau2016,Zeiher2016,Labuhn2016,Ebert2014}. While the lifetime is fundamentally limited by natural decay rates, an increase in Rabi frequency is mainly a technical challenge. 

\section{Experimental techniques}

Combining an optical lattice experiment with the excitation of Rydberg atoms introduces a new length scale in the system, but it also leads to challenges on the experimental side. It is desirable to prepare atom distributions with one atom per lattices site with spatial structures on the order of the blockade radius of the Rydberg atoms. Another difficulty are the extremely different energy scales of Rydberg excitation and lattice physics. Forces between the Rydberg atoms can be larger than all energy scales of the lattice and can lead to movement of the atoms before imaging.
In the following, we give a short overview of experimental techniques from preparation of ground state atoms to imaging of Rydberg atoms in an optical lattice.

\subsection{Preparation of atoms in optical lattice}

Densely filled systems with one atom per site can be prepared in optical lattices by driving the superfluid-Mott-insulator transition with ultracold bosonic atoms with repulsive interactions \cite{Greiner2002,BlochDalibardZwerger2008}. In this way more than $\SI{95}{\%}$ filling can be reached in the central region of the lattice.
Employing single-site addressing techniques, the atom distribution can be tailored to the needs \cite{Weitenberg2011,Preiss2015}.
The single-site addressing implemented in ref. \cite{Weitenberg2011} starts from atoms that are prepared spin-polarized in the optical lattice with one atom per site. Then a spin-selective optical light shift with the desired pattern is imaged onto the atoms. This will shift a microwave-transition between two hyperfine ground states for targeted atoms compared to the ones in the dark and allows for a selective microwave transfer via a narrow transition to another hyperfine spin state. 
In the end, the possible shapes are only limited by the size of the initial unit-filling region in the lattice and the resolution of the imprinted light pattern.  
 
Recently also other techniques were demonstrated to achieve a near-perfect array arrangement of atoms. Progress in the deterministic loading of microtraps allows for loading many traps with exactly one atom \cite{Barredo2016,Endres2016,Kim2016}. Compared to the lattice approach discussed above, the larger spacing between these traps can be an advantage for Rydberg experiments. Progress on Rydberg excitation in microtraps is reviewed in refs. \cite{Saffman2010,Browaeys2016}.

Another promising approach is to use atoms in an array of magnetic traps \cite{Leung2014,Wang2016,Whitlock2017} though the deterministic preparation of single atoms in these traps has not been demonstrated yet.

\subsection{Rydberg excitation and detection}

\begin{figure}
\includegraphics{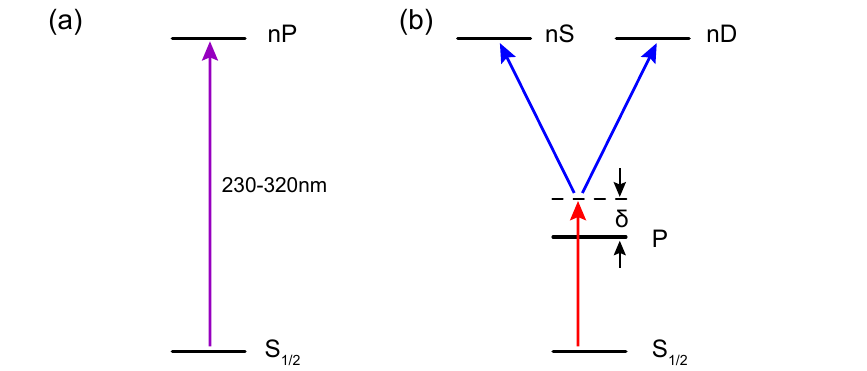}
\caption{Illustration of commonly used Rydberg excitation schemes for alkaline atoms. (a) Direct Rydberg excitation using an ultraviolet laser from the ground state. (b) Two-photon excitation via an intermediate P-state to nS or nD Rydberg states. Depending on the choice of the intermediate P-state the wavelength of the first transition can be longer than for the second transition or shorter (so-called inverted scheme). To allow for adiabatic elimination of the intermediate state a detuning $\delta$ is required. \label{fig:excitation_scheme}}
\end{figure}

The most common Rydberg excitation schemes are either a two-photon excitation to the Rydberg state or a direct excitation with a single laser from the ground state (Fig.~\ref{fig:excitation_scheme}). Both schemes have their advantages and drawbacks shortly discussed here. The two-photon scheme typically requires laser wavelengths that are well accessible with diode lasers or frequency-doubled diode lasers. In contrast, the direct excitation wavelengths are typically in the ultraviolet wavelength range and require more challenging laser systems. Another big advantage of the two-photon excitation scheme is the accessibility of both Rydberg S- and D-states for alkali atoms leading to more flexibility compared to the direct excitation using a single laser, which only gives access to P-states. Two-photon excitation is limited by the smallest of the two transition matrix elements of both transitions. For Rydberg excitation the limiting matrix element is the final one to the Rydberg state and for the two-photon excitation this matrix element is larger than the direct S-P matrix elements to Rydberg states of the same principal quantum numbers. But there are some drawbacks of the two-photon excitation, which are light shifts and scattering from the intermediate state. Scattering can be reduced by more intensity for the second excitation step or a narrow linewidth of the intermediate state (if available in the atom) and light shifts are in principle no problem as long as they are stable. This stability requirement can become challenging as it increases requirements for laser intensity stability tremendously. For the direct excitation light shifts are minimized. However, for experiments that use resonant excitation of Rydberg atoms and have enough intensity for the second excitation step available by either strong focussing or a lot of laser power, none of the schemes has big advantages over the other.

\begin{figure}
\includegraphics{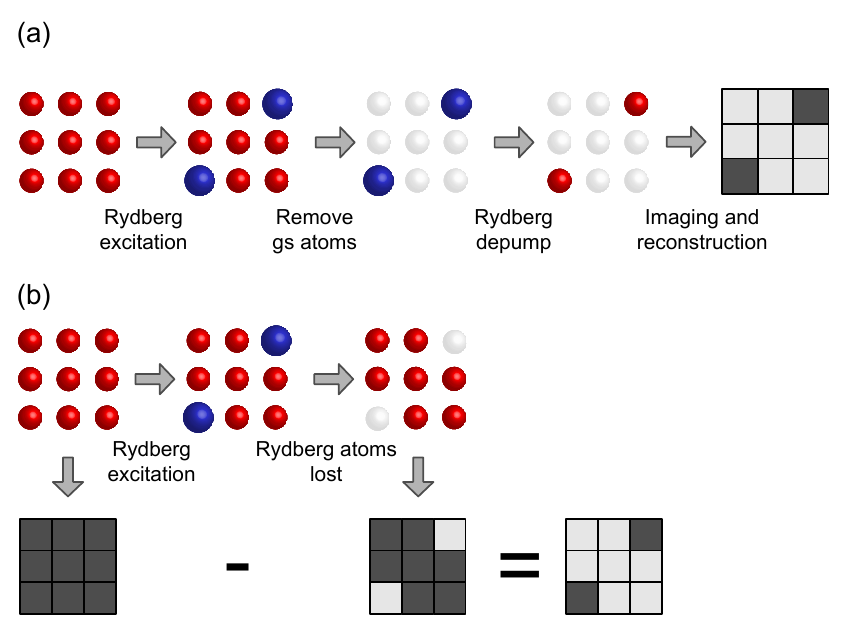}
\caption{Detection schemes of Rydberg atoms. (a) Detection in optical lattice as introduced in ref.~\cite{Schauss2012}. From left to right: Atoms in an initial configuration are coupled to a Rydberg state. Thereafter, ground state atoms are removed and in the following the Rydberg atoms are deexcited to the ground state. Finally, the remaining ground state atoms carry the information of the Rydberg atom positions and are imaged in a quantum gas microscope. The image is then digitized by applying a reconstruction algorithm \cite{Sherson2010}.
(b) Detection in optical tweezer arrays. Detection scheme as used in ref.~\cite{Labuhn2016}. From left to right: Atoms are prepared and imaged in the initial configuration and are then coupled to a Rydberg state. Then Rydberg atoms are lost as they are anti-trapped in the tweezers. Finally the remaining ground state atoms are imaged. The positions of the Rydberg atoms are obtained by subtracting the remaining atom configuration from the initial atom configuration.\label{fig:detection}}
\end{figure}

There are many ways Rydberg atoms were detected historically. The standard way is to apply electric fields and ionize the Rydberg atoms and guide the resulting ions and/or electrons to detectors \cite{Gallagher1994}. Gaining spatial resolution from this approach is possible but challenging \cite{Schwarzkopf2011,McQuillen2011,Dudin2012c,Lochead2013,Santra2015,Stecker2017}. The required in-vacuum components are also hard to combine with a high-resolution imaging and an optical lattice setup. Another approach is the detection of Rydberg atoms as loss, a technique that is often used in optical tweezer experiments \cite{Saffman2010,Browaeys2016} (Fig.~\ref{fig:detection}b). In an optical lattice this technique is on the one hand problematic as the Rydberg atoms are typically not lost with high fidelity and on the other hand it requires high fidelity imaging of the atom distribution before and after Rydberg excitation. This double imaging is challenging in a quantum gas microscope because the heating of the atoms needs to be kept low enough during the first imaging phase that they do not tunnel in the lattice during the experiment. Optimized Raman sideband cooling might allow for such a scheme in the future.
The requirement of double imaging can be avoided by detecting the Rydberg atoms themselves after de-exciting them in a controlled way to the ground state and blowing the other ground state atoms \cite{Schauss2012} (Fig. \ref{fig:detection}(a)). In this way spatial correlations between the Rydberg excitations can be extracted with high fidelity because features in spatial correlation functions are insensitive to uncorrelated loss of Rydberg atoms in the process.
Alternative ways to optically image Rydberg atoms are direct imaging on another transition available in alkaline-earth-like atoms \cite{McQuillen2013} or EIT-imaging \cite{Olmos2011,Guenter2012,Guenter2013}. For both of these techniques there are no major obstacles expected when extending them to imaging of single Rydberg atoms with single-site resolution in a lattice configuration. It is also possible to detect the photons from the decay of Rydberg atoms \cite{Dudin2012b,Goldschmidt2016} but it seems challenging to achieve site-resolved detection of single Rydberg atoms in this way.

\section{Limit of strong interactions - Isolated Superatoms}

The time-evolution in the Rydberg Hamiltonian leads in general to complicated entangled many-body states and is barely tractable numerically in mesoscopic systems. It is illustrative to first focus on a simple limit. One interesting case is the limit of dominating Rydberg-Rydberg interactions. In this setting, the blockade radius is larger than the diameter of the system and there can be only maximally one Rydberg excitation (Fig.~\ref{fig:superatom}(a)).
As a consequence, this system behaves as a collective two-level system with basis states of zero and one Rydberg excitation \cite{Jaksch2000,Lukin2001}. Experiments with two atoms in optical tweezers have first demonstrated this concept \cite{Urban2009,Gaetan2009}. Here we focus on a system with many ground state atoms. Such a two-level system offers promising perspectives for quantum computing as it allows for the storage of a qubit while being insensitive to atom loss \cite{Saffman2010,Saffman2016}. Moreover, gaining excellent control over such collective Rabi oscillations even constitutes one promising approach to implement quantum gates via collective encoding \cite{Saffman2008}.

\subsection{Superatoms as effective two-level system\label{sec:superatommodel}}
\begin{figure}
\includegraphics{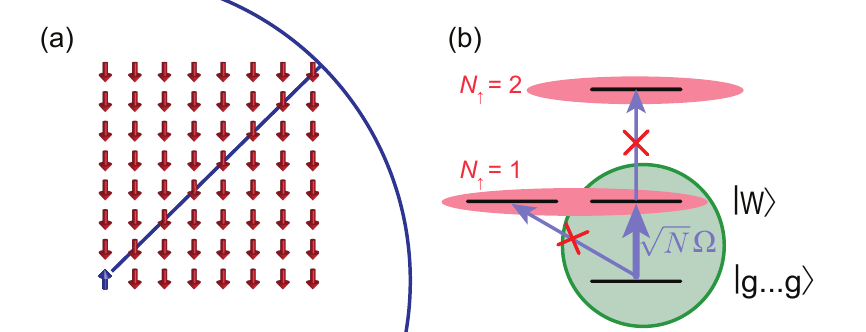}
\caption{Illustration of a superatom. (a) An array of spins with the blue circle showing the minimal blockade radius required to obtain a superatom in this system. (b) Level scheme of a superatom. The ground state is coupled with the collectively enhanced Rabi frequency $\sqrt{N}\Omega$ to the symmetric $\ket{W}$-state. The coupling to doubly-excited states and all other singly excited states is suppressed by detuning and symmetry, respectively. The green circle highlights the resulting effective two-level system. \label{fig:superatom}}
\end{figure}

The isolated superatom arises in a finite system in the limit of dominant interactions. Because there is only maximally one Rydberg atom in the whole system the state space is dramatically reduced and the full system can be described as a collective two-level system in the symmetric subspace of zero ($n_\uparrow=0$) and one ($n_\uparrow=1$) excitations (Fig.~\ref{fig:superatom}b). 
Here the ground state is the state with all $N$ atoms in the ground state $\ket{0}=\ket{\downarrow_1,\dots,\downarrow_N}$ and the excited state is the entangled first Dicke state $\ket{W}=\frac{1}{\sqrt{N}}\sum_{i=1}^{N}\ket{\downarrow_1,\dots,\uparrow_i,\dots,\downarrow_N}$, where $\downarrow_i$ and $\uparrow_i$ label the i-th atom in the ground or Rydberg state \cite{Dicke1954}. 
In the state $\ket{W}$ the single Rydberg excitation is symmetrically shared among all $N$ atoms under the assumption that both coupling and interaction are uniform. In this ideal situation, the $W$-state is the only state coupled by light from the ground state. 
With these definitions and assuming resonant coupling with $\Delta=0$, the Hamiltonian can be rewritten in the simple form $H=\hbar\sqrt{N}\Omega/2 \left( \ket{0}\bra{W}+\ket{W}\bra{0} \right)$. In this collective two-level system, the coupling is not given by the bare coupling $\Omega$ but by the symmetry-induced collectively enhanced coupling $\Omega_\text{coll} = \sqrt{N}\Omega$.

\subsection{Implementations}

\begin{figure*}
\includegraphics{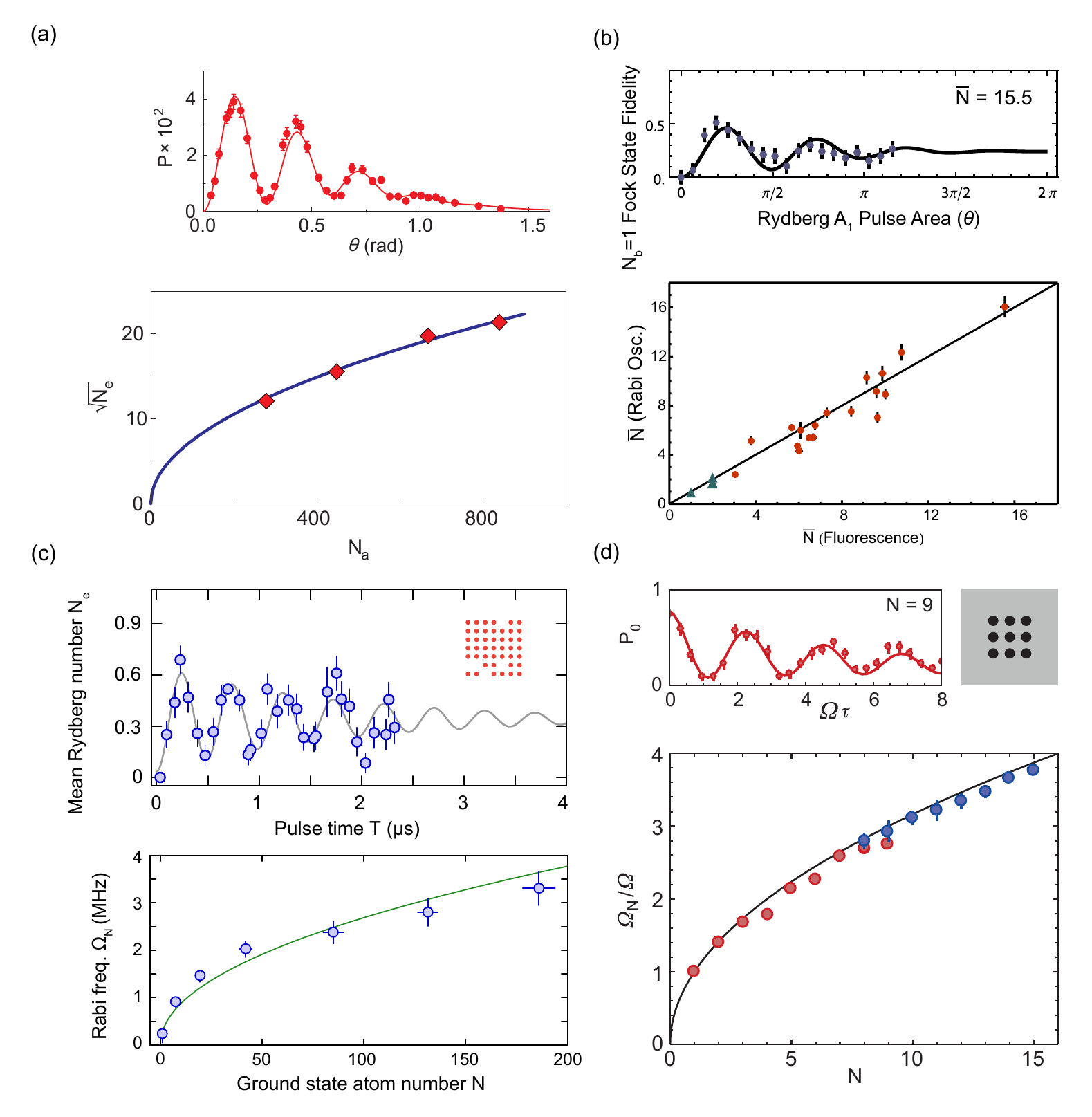}
\caption{Collective Rabi oscillations. Collective Rabi oscillation and $\sqrt{N}$-dependence of the collective Rabi frequency observed in various experiments with disordered atom positions (a-b) and atoms on a grid (c-d). (a) Collective Rabi oscillations detected via the emitted photons. Top, probability to detect a single photon versus single-atom Rabi angle $\theta$. Bottom, $\sqrt{N_e}$, determined from the collective enhancement of the Rabi frequency, as a function of the number of atoms $N_a$ measured via fluorescence imaging. Line is a fit ($0.74\sqrt{N_a}$). Reprinted figure from \cite{Dudin2012b}. (b) Collective Rabi oscillations with atoms in a single tweezer. Top, probability for detection of a single Rydberg atom versus single-atom Rabi angle $\theta$ for an average number of $\bar{N}=15.5$ atoms in the trap. Bottom, number of atoms inferred from collective enhancement of the Rabi oscillation as a function of the number of atoms measured via fluorescence. Exact initial atom numbers (triangles) or Poissonian atom number fluctuations in initial atom number (circles). Line, identity. Reprinted figure from \cite{Ebert2014}. (c) Collective Rabi oscillations with atoms in an optical lattice. Top, average number of detected Rydberg atoms as a function of the Rabi pulse time. Initial state is a $7\times7$-square of atoms in the lattice with average filling of 42(4). Inset, typical initial atom pattern. Bottom, fitted collective Rabi frequency $\Omega_N$ as a function of ground state atom number (blue circles) and theoretical prediction $\sqrt{N}\Omega$ (green line). Reprinted figure from  \cite{Zeiher2016}. (d) Collective Rabi oscillations in an array of optical tweezers with maximally one atom per trap. Top, probability to detect no Rydberg atom $P_0$ as a function Rabi pulse area $\Omega\tau$. The initial configuration of the preselected nine atom configuration is shown on the right. Bottom, collective enhancement of the Rabi oscillation $\Omega_N/\Omega$ versus atom number $N$. Fully loaded arrays (red dots) or initial samples with fixed atom number but random distribution (blue dots). Black line, theoretical prediction $\sqrt{N}$. Reprinted figure from \cite{Labuhn2016}.  \label{fig:collective_rabi}}
\end{figure*}

The first implementations of isolated superatoms were created to demonstrate a two-qubit gate with Rydberg atoms \cite{Gaetan2009,Urban2009,Isenhower2010,Wilk2010,Hankin2014}. These experiments implement the minimal superatom with exactly two atoms and observed the $\sqrt{2}$ enhanced collective Rabi oscillation  \cite{Saffman2010,Browaeys2016}. 

Other implementations were driven by the idea to study single photon, single electron or single ion sources \cite{Dudin2012,Li2013,Balewski2013,Weber2015} and exploit the fact that the superatom can have maximally one Rydberg atom. For these purposes the coherence of the superatom is not required and in some cases even disadvantageous. In contrast, for quantum computing applications the coherence of the superatoms is essential.  
Collective Rabi oscillations were observed in larger ensembles \cite{Dudin2012b,ParisMandoki2017} and for few atoms with preselected atom number \cite{Ebert2014} (Fig.~\ref{fig:collective_rabi}). The spatial ordering of the atoms is not important for superatoms as long as all atoms are within the blockade radius and ground state-Rydberg atom interactions can be neglected.

Recently systems with many atoms allowed to confirm the $\sqrt{N}$ scaling of the collective Rabi frequency over about two orders of magnitude \cite{Zeiher2015,Labuhn2016}. Here the initial atom number was known with sub-Poissonian fluctuations by either employing the Mott insulator transition in an optical lattice, or measuring the number of atoms before Rydberg excitation.

The entangled nature of the excited state in the superatom has been also inferred \cite{Zeiher2015,Ebert2015}. The underlying idea is to determine the population in the coherent superatom sector by determining the amplitude of the Rabi oscillation. Based on this quantity an entanglement witness can be constructed to show $k$-particle entanglement for a significant part of the ground state atom number. But the predicted $W$-state has not been directly confirmed via quantum state tomography yet.

\subsection{Challenges}

Collective Rabi oscillations have been observed with fully-blockaded Rydberg ensembles in various experiments. The coherence time of the collective Rabi oscillation in all of them is on the order of a few oscillations and shorter than expected and the different influences of some contributing effects are not fully understood. Next to effects caused by technical noise that can scale with the collective rather than the bare Rabi frequency, there are other more complicated dephasing processes. One is for example the coupling of the excited state to the set of dark states with one excitation that grows linearly with the number of spins. Another is the off-resonant coupling to the doubly excited states, discussed in the next section. There are also contributions from black body radiation and imperfect blockade due to coupling to molecular states. A more systematic study of these effects is necessary to understand the dephasing effects in more detail and improve the coherence times.
Another technical difficulty is the preparation of a deterministic atom number in the fully-blockaded ensemble. Variations in atom number are hard to distinguish from dephasing in the measurements. The recent work on reduction of temperature and noise in quantum gas microscopes will also lead to improvements here. But also new techniques for deterministic preparation of arbitrary atom configurations have been developed for tweezer arrays \cite{Barredo2016,Endres2016,Kim2016}.

\section{Beyond isolated superatoms}

In experiments the Rabi coupling in the superatom is typically not fully negligible compared to the interaction shift caused by two Rydberg excitations in the system. As the coherent collective Rabi oscillation will off-resonantly couple to a large two-excitation space growing quadratically with the number of spins, even small couplings lead to relevant dephasing. Understanding of these effects can be gained by breaking the full blockade condition on purpose by spreading the atoms out over an area with a diameter close to or larger than the blockade radius. 
This regime has been explored experimentally with systems on the order of the blockade radius \cite{Barredo2014,Zeiher2015}.
Even continuum systems much larger than the blockade radius still show a collectively enhanced Rabi frequency \cite{Heidemann2007,Stanojevic2009,Viteau2011a,Valado2013}.

\begin{figure}
\includegraphics{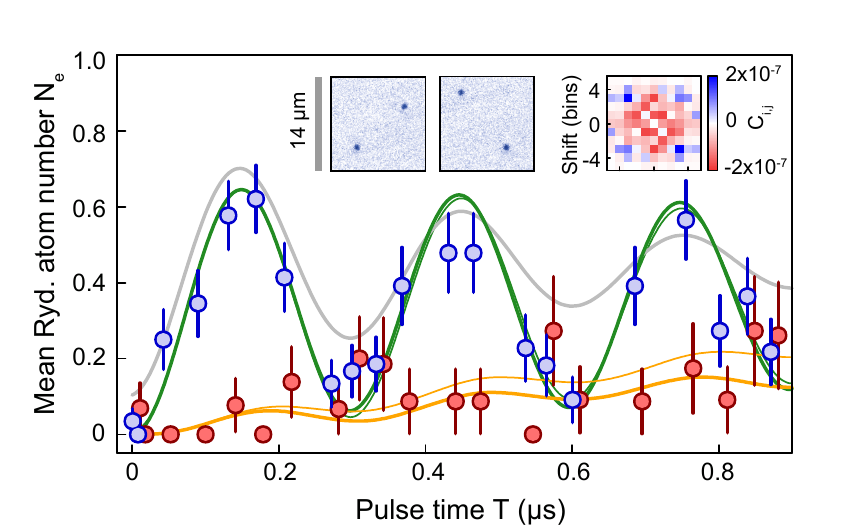}
\caption{Breaking the superatom. Collective Rabi oscillation with coupling to a two-excitation state in a 15x15 atom initial lattice. Experimental single atom and two-atom events are shown (blue and red points) together with two theoretical predictions. One is a calculation based on a reduced basis set with $R_c = 8$ (thick lines), the other is an approximation assuming a perfect Dicke state in the singly excited subspace and only coupling to symmetric two-atom states (thin lines, see text). Both take into account the experimental initial atom configurations with an average filling of $0.82$ as well as the detection efficiency of $\eta=0.68$. Gray line is a fit to the experimental average excitation number. The insets show two experimental two-excitation pictures (left) and the correlation function between two-atom events (right).
Adapted from ref.~\cite{Zeiher2015}.
\label{fig:beyondsuperatom}}
\end{figure}

\begin{figure}
\includegraphics[width=0.3\textwidth]{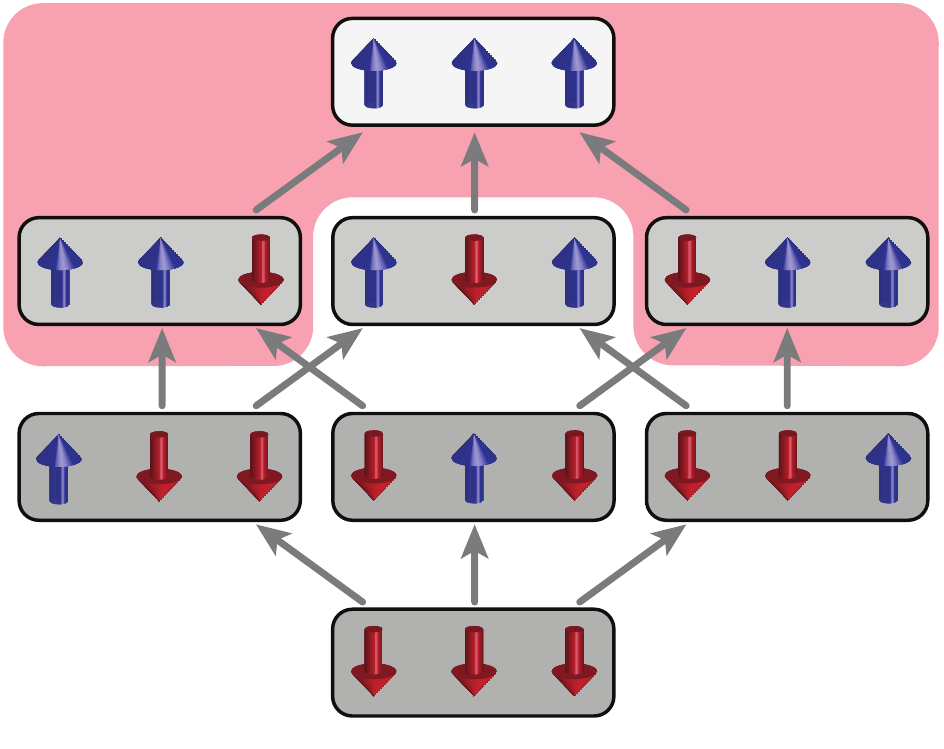}
\caption{Minimal example of three spins. The configurations are coupled by spin-flips with coupling $\Omega$ (gray arrows). The interaction energy of each state is indicated by the brightness of the gray background. When selecting states with a minimal distance between spin-$\uparrow$ of $R_c = 1.5$ all states in the red shaded area are neglected. While the coupling from the ground state $\ket{\downarrow\downarrow\downarrow}$ to the state states with $n_\uparrow=1$ is spatially homogeneous, the coupling of the singly excited states to the state with $n_\uparrow=2$ is spatially inhomogeneous.
\label{fig:states}}
\end{figure}

To illustrate the effects of breaking the full blockade condition slightly in a superatom we consider a simple model. We imagine a square of atoms in a lattice where the diagonal is comparable to the blockade radius. In such a setup the first pairs of Rydberg atoms are expected along the diagonal of the system since these doubly excited states exhibit the smallest detuning. To describe this system we can extend the simple two-level model discussed above and add a state with two excitations aligned along the diagonals $\ket{\diagup}$ and $\ket{\diagdown}$. Only the symmetric superposition $\ket{2} = \frac{1}{\sqrt{2}} \left(\ket{\diagup}+\ket{\diagdown}\right)$ of these states is coupled with a Rabi frequency $\bra{W}\hat{\Omega}\ket{2} = 4/\sqrt{2 N} \Omega$ to the $W$-state ($\bra{\downarrow}\hat{\Omega}\ket{\uparrow} = \Omega$). More generally one finds a coupling of $C/\sqrt{N M} \Omega$ with $C$ the number of coupled pairs of states with $\Omega$ between a $N$-fold degenerate singly excited state and a $M$-fold degenerate doubly excited state \cite{Pohl2010a,Schachenmayer2010,VanBijnen2011,Petrosyan2016}. We end up with a Hamiltonian with the states $\{ \ket{0}, \ket{W},\ket{2} \}$:

\begin{equation}
H = \hbar \Omega \left( \begin{array}{ccc} 0 & \frac{\sqrt{N}}{2} & 0 \\  \frac{\sqrt{N}}{2} & 0 &  \frac{C}{2\sqrt{N M}} \\ 0 & \frac{C}{2\sqrt{N M}} & -\frac{C_6}{d^6}  \end{array} \right) \; , 
\end{equation}
where $d$ is the distance between the two atoms in the doubly excited state.

Surprisingly, this model fails to predict the coupling to doubly excited states quantitatively. The reason is that many two-excitation states with different energy have a considerable influence. Taking all of them into account in the same way as above leads already to a reasonable approximation. For that purpose we introduce one additional state for each appearing distance between two atoms. As simplification we couple the $W$-state equally to each of these states which consist of a symmetric superposition of all doubly excited states with same energy. The result of this approximation is shown together with experimental data in Fig.~\ref{fig:beyondsuperatom}. But there is another important effect which is the spatially inhomogeneous coupling of the singly excited states to the doubly excited ones. This leads to a deviation from the $W$-state in the singly excited subspace (Fig.~\ref{fig:states}). It manifests itself as a spatially dependent variation of the amplitude factors around $1/\sqrt{N}$. This spatial dependence can be taken into account in the model by treating all states with $n_\uparrow=1$ separately and not only as one $W$-state. Thereby one can describe the dynamics on a nearly blockaded $15\times15$ square with a $164\times164$ Hamiltonian as well as the reduced basis set calculation that takes into account many more states (Fig.~\ref{fig:beyondsuperatom}). These approximations are, of course, only reasonable for short evolution times in the special case of weak breaking of the full blockade condition where the interaction energy of the lowest two-excitations states is on the order of the Rabi-frequency. For long evolution times weak couplings to more states have to be considered.
The reduced basis set calculation considers only states that have a minimal distance between excitations larger than a critical distance $R_c$ \cite{Pohl2010a}. This idea relies on the observation that the states with closer excitations have extremely high energy and are typically far off-resonant and not coupled. Care has to be taken for anti-blockade configurations \cite{Ates2007,Amthor2010}. In any case the validity of this approximation can be checked by looking at the convergence of the solution for $R_c \rightarrow 0$.


\section{Sudden Quench in the Rydberg Hamiltonian}

For systems larger than the blockade radius the approximations discussed in the previous sections fail and in general the full Hamiltonian needs to be considered. The most commonly used technique to investigate the Rydberg Hamiltonian is a sudden switch-on of the coupling to the Rydberg state for a certain time. This corresponds to a quench of the transverse field in the spin Hamiltonian. The sudden quench of the coupling from the ground state with $n_\uparrow=0$ leads to dynamics because it is not an eigenstate of the Hamiltonian. The initial state is projected to the basis states leading to a superposition with wide spread of energies that show different time evolution. However, experiments show that for atom numbers larger than two, the system ends up in a quasi steady-state on a timescale on the order of $2\pi/\Omega_\text{coll}$. In the following, we look in more detail at the dynamics and then at the properties of the quasi-steady state.

\subsection{Excitation dynamics} 

Here we consider the following excitation sequence. Initially, all atoms are spin-polarized in the ground state and at time $t = 0$ the coupling $\Omega$ to the Rydberg state is switched on resonantly (detuning $\Delta = 0$). After an evolution time, the system is detected at a time $t = t_0$. This evolution is equivalent to a quench of the transverse magnetic field in a long-range interacting Ising Hamiltonian.
The excitation dynamics in the Rydberg Hamiltonian is quite complicated, especially in the general case where the system is neither fully-blockaded nor in the low-excitation limit. The van-der-Waals interaction leads to a huge variety of energy scales that cause interaction-induced dephasing of the system. The coherence is preserved for longer than Rabi oscillations can be observed in the system average. But  coherence is hard to measure as it becomes only accessible in experiments that are sensitive to the phases \cite{Raitzsch2008,Younge2009,Schauss2012}. All density-related observables seem to show a steady-state behaviour that has been investigated \cite{Stanojevic2009,Olmos2012} and observed in many experiments \cite{Liebisch2005,Deiglmayr2006,Liebisch2007,Heidemann2007,Johnson2008,Reetz-Lamour2008,Low2009,Viteau2012,Schempp2014,Malossi2014,Labuhn2016}. For resonant driving of the Rydberg transition, this state is reached in a time on the order of $\frac{2\pi}{\sqrt{N_b}\Omega}$ \cite{Stanojevic2009,Olmos2012,Ates2012b,Ates2012d}, where $N_b$ is the number of atoms per blockade sphere. Theory calculations show that the observation of quasi steady states can be explained by fully coherent dynamics and does not rely on external decoherence sources.
The time evolution can be seen intuitively as a superposition of many Rabi oscillations from the ground state to all singly excited states which are then further coupled to doubly excited states and so on. For small system sizes, the number of excited states is limited and the relevant dynamics of the system can be understood by the approximations discussed in the previous section. 
The superatom idea can be extended to systems with larger number of excitations leading to a picture of a dense packing of Rydberg blockade spheres for large excitation numbers \cite{Ji2011,Ates2012b}. In this regime it becomes hard to approximate the full dynamics but it provides access to the excitation statistics of the steady-state. The typical timescales of the excitation dynamics are still given by a collectively enhanced Rabi frequency where the effective number of atoms participating is roughly $N_b$ \cite{Heidemann2007,Younge2009,Stanojevic2009,Viteau2011}.

\subsection{Spatial correlations}

Spatially resolved detection of the Rydberg atoms allows to directly measure correlation functions and thereby also the blockade radius. The discussion of the excitation dynamics in the last section shows that the spatial correlation function evolves to a steady-state.  In this regime the Rydberg atoms arrange randomly but are subject to the blockade constraint and thereby show liquid-like distance correlations. 
There has been a variety of theoretical work on the steady state patterns expected in the Rydberg gas \cite{Olmos2009a,Hoening2012,Gaerttner2013}. It has been argued that high density initial states entropically favour ordered Rydberg excitations \cite{Ates2012b}. Another view on the ordering is that the spin-$\uparrow$-distribution can be modelled in driven-dissipative steady-state by particles interacting with logarithmic potentials \cite{Lechner2015c}.

One signature for the appearance of correlations is the excitation statistics of Rydberg atoms in the system. It has been observed experimentally that the Rydberg numbers are sub-Poissonian \cite{Liebisch2005,Liebisch2007,Viteau2012,Schempp2014,Malossi2014}.

\begin{figure}
\centering
\includegraphics{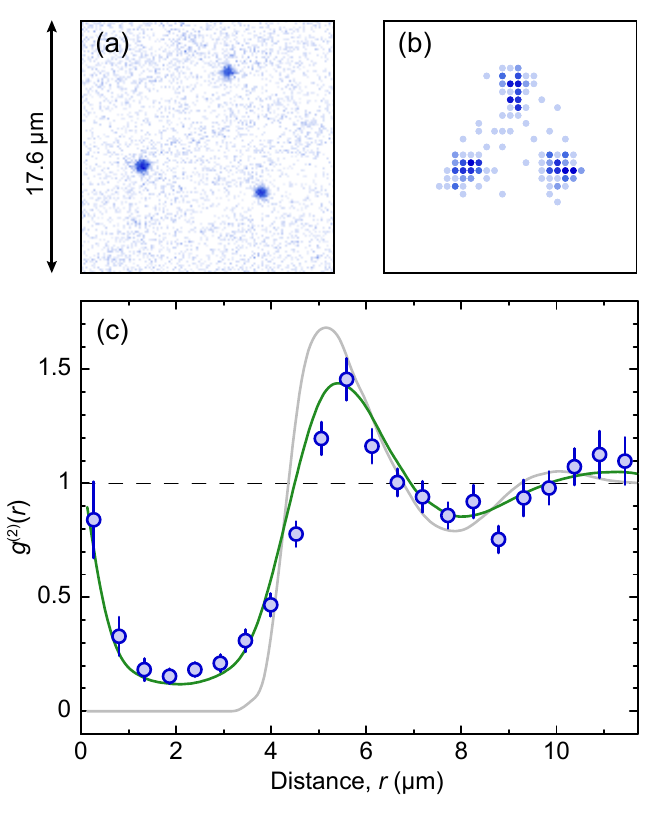}
\caption{Pair correlation function in a 2D system. (a) Single shot fluorescence image of three Rydberg atoms. (b) Aligned pictures from many repetitions illustrating the correlations (see ref.~\cite{Schauss2012} for details) (c) Radial correlation function. The blockade effect results in a strong suppression of the probability to find two excitations separated by a distance less than the blockade radius $\Rbk = \SI{4.9 +- 0.1}{\micro m}$. The spacing of atoms in the lattice is $\alat=\SI{532}{nm}$. The experimental data (blue circles) are compared to the ideal theoretical prediction (grey line) and taking into account the independently characterized imperfections of the detection method (green line). The rise of the correlation function below $\SI{1}{\micro m}$ is an imaging artifact (see text). The dashed line marks the value of \gtwo in the absence of correlations. The error bars represent the standard error of the mean of $\gtwo(r)$. Adapted from ref.~\cite{Schauss2012}.\label{fig:spatial_correlation_munich}}
\end{figure}

\begin{figure}
\centering
\includegraphics{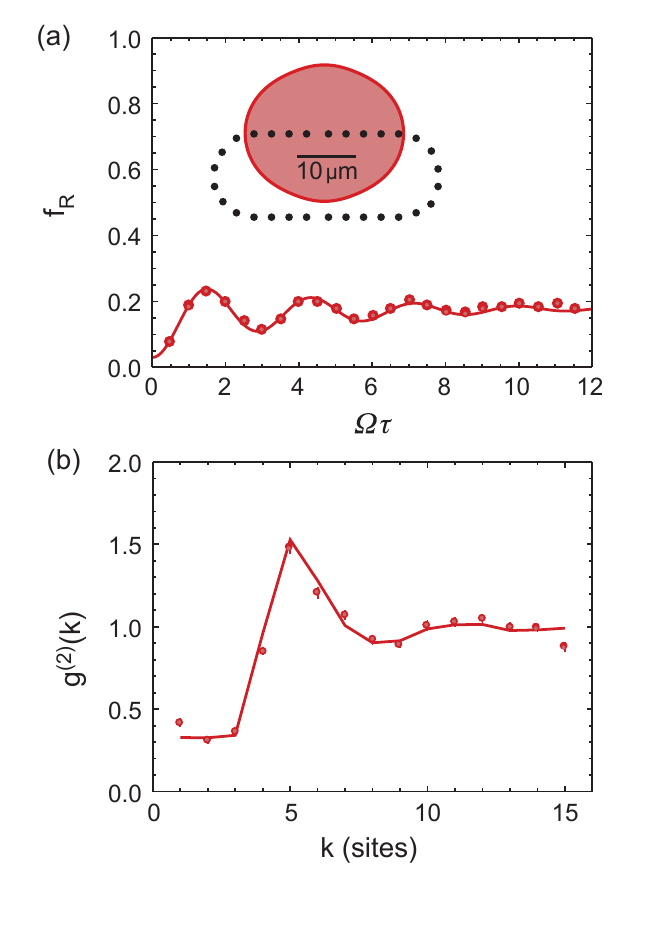}
\caption{Pair correlation function in a 1D system. (a) Rabi oscillation and dephasing induced by the beating of the many eigenfrequencies of the interacting ensemble of atoms. The Rydberg population $f_R$ is shown for experiment (points) and theory (line). Inset shows the configuration of optical tweezers and the red ellipse illustrates the size of the anisotropic blockade radius. (b) Pair correlation function at $\Omega\tau \approx 2$ from (a) (points) with theory model (line). Reprinted from \cite{Labuhn2016}. \label{fig:spatial_correlation_paris}}
\end{figure}

A more direct way to measure spatial correlations is to extract them from high-resolution images. 
These correlations have been studied by ionizing the Rydberg atoms \cite{Schwarzkopf2011,Schwarzkopf2013} and in-situ with optical lattices \cite{Schauss2012} (Fig.~\ref{fig:spatial_correlation_munich}) and optical micro-traps \cite{Labuhn2016} (Fig.~\ref{fig:spatial_correlation_paris}).
The spatial correlations have been characterized by the pair correlation function


    \begin{equation}
    \label{eq:g2r}
    \gtwo(r) = \frac{\sum_{{\vec i} \neq {\vec j}} \delta_{r, r_{\vec i\vec j}} \, \langle
      \hat\sigma^{(\vec i)}_{ee} \hat\sigma^{(\vec j)}_{ee} \rangle} {\sum_{{\vec i} \neq {\vec j}}
      \delta_{r, r_{\vec i\vec j}} \, \langle \hat\sigma^{(\vec i)}_{ee} \rangle \langle
      \hat\sigma^{(\vec j)}_{ee} \rangle} \; .
  \end{equation}

  It measures the joint probability of two excitations at a distance $r$. Here
  $\delta_{r, r_{\vec i\vec j}}$ is the Kronecker symbol that restricts the sum to sites $(\vec i,
 \vec j)$ for which $r_{\vec i \vec j} = r$.
 The basis for the calculation of this correlation function is a set of data of positions of all spin-$\uparrow$ atoms in many realizations of the experiment (Fig.~\ref{fig:spatial_correlation_munich}(a)). A first idea of the correlations can be gained by aligning center of mass and angle of events with same spin-$\uparrow$ number (Fig.~\ref{fig:spatial_correlation_munich}b). For the calculation of the correlation function all realizations independent of the number of spin-$\uparrow$ atoms are included (Fig.~\ref{fig:spatial_correlation_munich}c). This correlation function shows the expected features from theory \cite{Gaerttner2013,Petrosyan2013a,Petrosyan2013b,Lechner2015c}, which are the blockade effect for small distances and a peak slightly beyond the blockade radius. Imaging imperfections lead to non-vanishing correlations below the blockade radius and a slight smearing out of the edge. The sharp rise at very short distance is caused by tunnelling during imaging that leads to an atom to be falsely detected as two atoms with probability of about $\SI{1}{\percent}$ \cite{Schauss2012}. A signal below the blockade radius could in principle also arise from pair excitations \cite{Derevianko2015} but it is unlikely that pairs excited in that way can be detected in the applied imaging technique considering the forces acting between the two atoms in such a pair. Imperfect blockade is typically rather caused by imaging imperfections that lead to detection of Rydberg atoms in places where they not have been during excitation. This can either happen by detecting a ground state atom wrongly as Rydberg atom or due to movement of the Rydberg atoms during the imaging sequence. 
 The corresponding pair correlation function in a 1D system has been measured with an array of equidistant atoms in microtraps on a larger spatial scale but the correlation function still shows qualitatively the same features (Fig.~\ref{fig:spatial_correlation_paris}). In the same study also the influence of anisotropic interactions on the 2D correlation function has been observed \cite{Labuhn2016}.

\section{Adiabatic preparation of Rydberg crystals}

\begin{figure}
\includegraphics{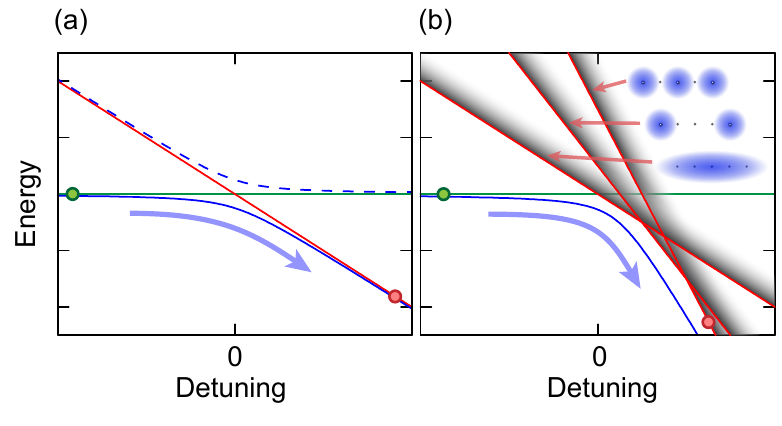}
\caption{Landau-Zener analogy between two-level and Rydberg many-body system. (a) Landau-Zener sweep in a two-level system. The system starts in the ground state (green circle) and via the sweep over the resonance at $\Delta=0$ it ends up in the excited state (red circle). Excited state and ground state for $\Omega=0$ are shown in green and red, respectively. The adiabatically connecting state is shown in blue. (b) Illustration of adiabatic sweep in many-body Rydberg setting. The  energy of ground state with $n_\uparrow=0$ is shown $\Omega=0$ as green line. The are several manifolds of excited states each characterized by the number of excitations (red lines). For each manifold there is a lowest energy state (excitation pattern illustrated by insets) and a variety of excited states indicated by the shading. As before the system starts out in the ground state (green circle) and evolves via the blue path to the excited state (red circle). Compared to (a) not only more state crossings need to be adiabatically traversed but also the switch-off of the Rabi frequency becomes relevant (see text).\label{fig:landau_zener_analogy}}
\end{figure}

The sudden switching of the coupling to the Rydberg state and the resulting time evolution discussed in the previous section poses the question if it is possible to perform adiabatic sweeps into collective Rydberg states and thereby deterministically excite a certain number of Rydberg excitations in the system. These states show interesting properties, such as crystalline ordering of the Rydberg excitations at low energy \cite{Weimer2008,Sela2011,Lesanovsky2011,Ji2011}. 
Adiabatic preparation has been proposed to deterministically prepare these states \cite{Pohl2010a,Schachenmayer2010,VanBijnen2011,Petrosyan2016}.
In the following we first discuss the general adiabatic preparation scheme in a many-body system and then the details of an implementation to prepare ordered Rydberg many-body states.

\subsection{Adiabatic preparation in a many-body system}

The adiabatic preparation in the many-body system follows the basic scheme of the Landau-Zener sweep in a two-level system \cite{Zener1932} (Fig.~\ref{fig:landau_zener_analogy}(a)). In both cases the underlying idea is to use a Rabi coupling to open a spectral gap, which allows to connect initial and final state by an adiabatic path. This path can be followed by changing the detuning and the coupling with time. Compared to the two-level system, the interacting many-body system has more energy scales which leads to a breakdown of the Landau-Zener picture. In the end the main problem is the appearance of the new interaction gaps in the many-body Hamiltonian (Fig.~\ref{fig:landau_zener_analogy}(b)).
The smallest gap in the process is not only determined by the Rabi frequency any more but by interactions \cite{Brierley2012}. So the simple recipe of the Landau-Zener sweep that an increase of Rabi frequency always improves the transfer is not true any more. In the many-body case one either has to adapt the duration of the adiabatic sweep to be slow compared to the gaps, enlarge the gaps or reduce the number state crossings. Larger gaps can be mainly achieved by decreasing the number of atoms, decreasing the distance between Rydberg atoms in the crystal or by  optimizing the path in $(\Omega,\Delta)$-space to avoid regions with small gaps \cite{Pohl2010a}. The number state crossings depends on the spatial configuration of the ground state atom distribution and is, for example, much smaller in 1D than in 2D.

\subsection{Crystallization}

\begin{figure}
\includegraphics[width=0.5\textwidth]{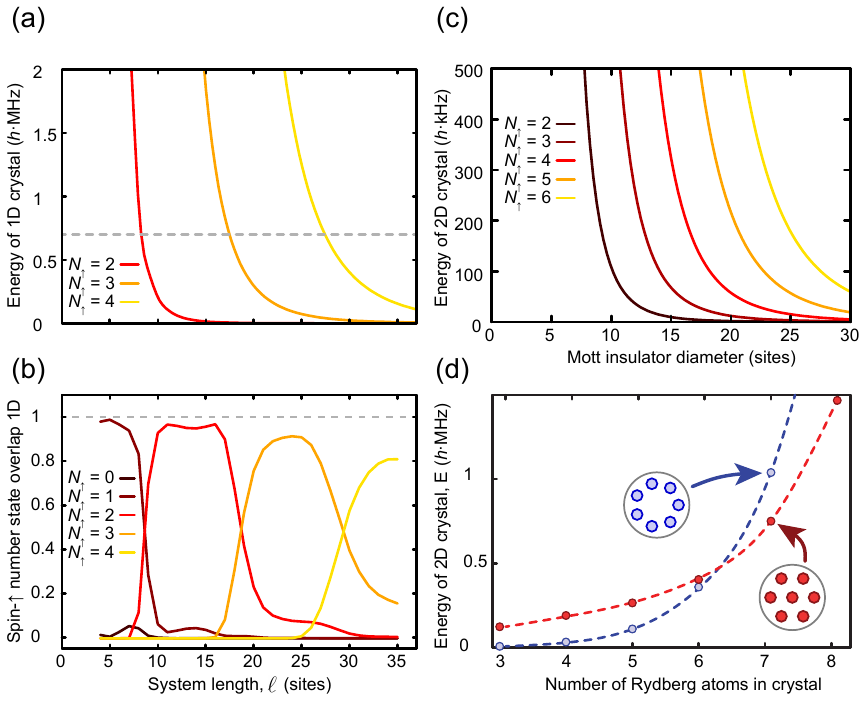}
\caption{Classical crystal interaction energies and relation to the adiabatic sweep. All calculations for the repulsively interacting 43S state of \Rb with $C_6 = -2.45$\,GHz$\cdot$\textmu m$^6$ and $\alat = \SI{532}{nm}$.
	(a) Crystal interaction energies in 1D of the classical linear crystalline configurations with $2$ to $4$ spin-$\uparrow$ atoms (red to yellow). The horizontal grey line marks the final detuning of the sweep as the most relevant energy scale. The positions of the crystal interaction energies crossing the final detuning line determine approximately at which length $\ell$ the spin-$\uparrow$ number after the sweep steps up by one.
	(b) Calculated dependence of the population of many-body states with fixed spin-$\uparrow$ atom number after the sweep on the system length. The calculation was done for an ideal $3\times\ell$ system with unity filling. Curves for the number of spin-$\uparrow$ atoms $N_\uparrow=0\ldots 4$ are shown. For $N_\uparrow\ge 4$ the efficiency of the preparation drops significantly due to smaller gaps on the trajectory of the sweep. 
  (c) Interaction energy of crystalline configurations in 2D neglecting lattice structure. Theoretical interaction energy of the crystal configurations with 2 to 6 spin-$\uparrow$ atoms (dark red to yellow) on the border of a disc-shaped sample.
(d) Illustration of the energy change from a ring-like crystal (blue circles) to a crystal with central atom (red circles) neglecting lattice structure. For up to six spin-$\uparrow$ atoms the ring crystal has lower energy, starting from seven the other configuration. Insets show the crystal configurations for seven spin-$\uparrow$ atoms.
\label{fig:crystal_energy}}
\end{figure}

\begin{figure*}
\includegraphics{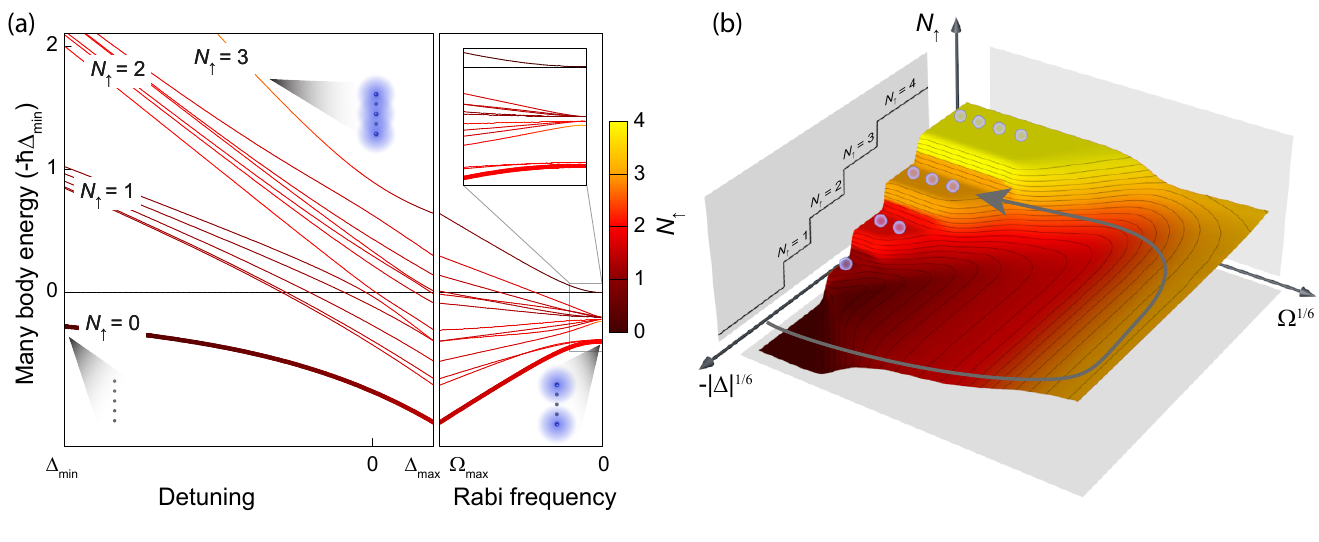}
	\caption{Many-body adiabatic sweep. (a) Illustration of a realistic adiabatic sweep in the simplified case of a five-atom system. In the left panel first only the detuning $\Delta(t)$ is changed, afterwards in the right panel only the coupling strength $\Omega(t)$. First, the detuning is changed from $\Delta_\text{min}$ to $\Delta_\text{max}$ at constant Rabi frequency $\Omega_\text{max}$, with $\Delta_\text{max}$ chosen to prepare $N_\uparrow=2$ (left panel).
    Subsequently, the Rabi frequency is reduced linearly from $\Omega_\text{max}$ to
    $0$ (right panel). The inset is a zoom into the end of the sweep focussing on the
    shrinking gap between the energy levels. The colour of each line indicates
    the mean number of spin-$\uparrow$ atoms in the many-body state. For large
    negative detuning the four different manifolds correspond to the
    crystalline states with fixed magnetization given by $N_\uparrow$ as
    indicated in the figure. In three limiting cases in which the states become
    classical the spatial distribution is shown (blue circles:
    Rydberg atoms, grey circles: ground state atoms).
	(b) Phase diagram in $(\Omega,\Delta)$-space calculated for $N=7$ atoms. The colour scale indicates the number of spin-$\uparrow$ atoms $N_\uparrow$ in the many-body ground state of a one-dimensional system. $N_\uparrow$ is also visualized in the crystalline phase by the small spheres. The detuning $\Delta$ and Rabi frequency $\Omega$ axes are rescaled by their sixth root. 
	Adapted from ref.~\cite{Schauss2015}. \label{fig:landau_zener_real}}
\end{figure*}

Adiabatic preparation techniques have been proposed to deterministically prepare Rydberg crystals \cite{Pohl2010a,Schachenmayer2010,VanBijnen2011,Petrosyan2016}. For an experimental implementation of these proposals it is important to first look at the energy scales in the system that determine the timescale for the preparation.
As the time for Rydberg experiments is typically limited to about \SI{10}{\micro\second} due to mechanical motion and lifetime of the Rydberg atoms, this imposes strict limits on the set of systems with large enough gaps to make the adiabatic preparation realistic. In contrast to adiabatic passage in two-level systems the transfer here is not limited by Rabi frequency but by the interaction gaps. Experimentally the preparation was demonstrated in one-dimensional systems with up to about 20 ground state atoms and up to three Rydberg excitations \cite{Schauss2015} (Fig.~\ref{fig:crystal_energy}(a-b)), which is an implementation in an optical lattice with spacing $\alat = \SI{532}{nm}$ using the rubidium Rydberg state 43S. The spin-$\uparrow$ number in the system changes by one when increasing the length by about 10 sites. Therefore a 1D system with length fluctuations much smaller than that is required. Average fluctuation of less than one site can be achieved using single-site addressing in a Mott insulator \cite{Weitenberg2011}. Experimentally, the number of Rydberg excitations is limited here to about three due to breakdown of adiabaticity for more than about 25 ground state atoms. Supporting theory calculations for the experimental parameters show that there is essentially no parameter space where one can prepare four excitations but not three for these experimental parameters (Fig.~\ref{fig:crystal_energy}(c)). Crystals with more excitations are realizable in an optimized setup with larger interaction gaps, for example by reducing the blockade radius, or by coupling to Rydberg states with longer lifetime.

For the experimental implementation an optimized time-dependence of $\Omega(t)$ and $\Delta(t)$ needs to be determined which is done via numerical optimization.  Typically, the sweep starts with a rise of the Rabi frequency followed by a sweep of the detuning over the resonance. This part can be relatively fast if sufficiently high Rabi frequency can be reached in the experiment. In the end the Rabi frequency needs to be reduced to reach the final state at $\Omega=0$. The timescale of this part is limited by the interaction gaps in the system. In Fig.~\ref{fig:landau_zener_real}(a) the spectrum during the sweep is shown for a simplified small system. The gap in the many-body spectrum during the adiabatic sweep is comparably large during the first half of the sweep. In the end during the reduction of the Rabi frequency the gap becomes very small. This shows the importance of a soft switch-off of the Rabi frequency in the experiment.
The approximate path of the adiabatic sweep through the $(\Omega/\Delta)$-space together with the average number of spin-$\uparrow$ atoms along the path is illustrated in Fig.~\ref{fig:landau_zener_real}(b).
The sweep starts at $\Delta<0$ in the classical limit ($\Omega=0$) where the system contains no spin-$\uparrow$ atoms. The target state is one of the classical crystalline states at $\Delta>0$. There the many-body ground state corresponds to crystalline states with vanishing
fluctuations in the total magnetization $M=2 N_\uparrow - N$, which, for fixed total atom number $N$, is determined by the
spin-$\uparrow$ component $N_\uparrow=\sum_{\vec i} \langle \hat{n}^{(\vec
i)}_\uparrow\rangle$. In a one-dimensional chain of $\ell\gg N_\uparrow$ lattice sites, the number of spin-$\uparrow$
atoms increases by one at the critical detunings $\ell^6\hbar\Delta_{\rm
  c}\approx 7 \abs{C_6} N_\uparrow^6 / \alat^6$ separating successive crystal
states with a lattice spacing $\alat\ell/(N_\uparrow-1)$~\cite{Pohl2010a}. The laser coupling introduces quantum fluctuations that can destroy the crystalline order~\cite{Weimer2008,Weimer2010b,Sela2011,Lesanovsky2011}. 

\begin{figure}
\includegraphics{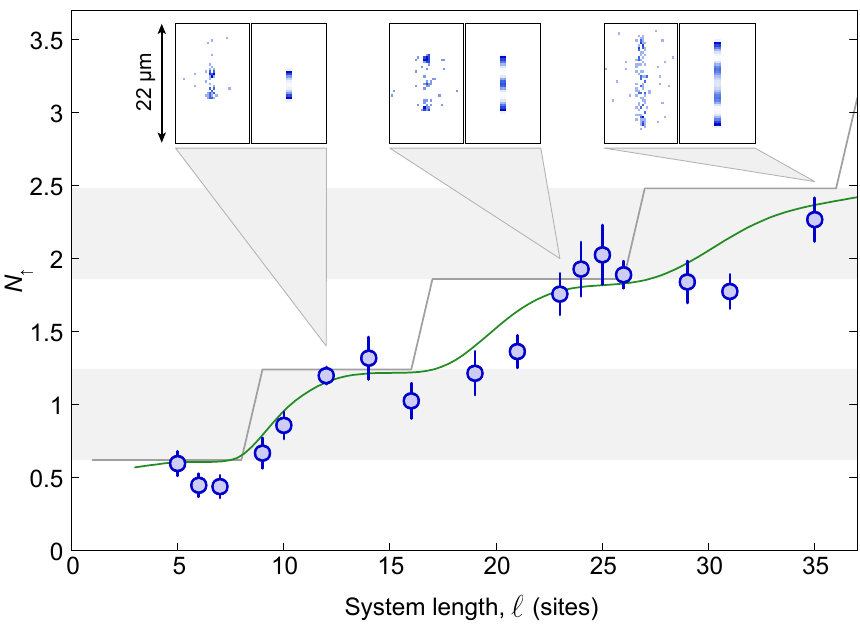}
\caption{Crystallization in 1D. The staircase of excitations is shown by the mean spin-$\uparrow$ number $N_\uparrow$ versus system length $\ell$ for a quasi-1D system. The green line is the theoretical prediction for the experimental sweep and the grey line the classical or fully-adiabatic expectation. Insets show the measured spatial distribution of the magnetization (left) and corresponding theory (right) for indicated system lengths. The brightness (light to dark) indicates the normalized number of
    spin-$\uparrow$ atoms. Adapted from ref.~\cite{Schauss2015}. \label{fig:staircase}}
\end{figure}

One clear signature for the successful implementation of the adiabatic preparation is a staircase of the number of spin-$\uparrow$ atoms versus length of the system (Fig.~\ref{fig:staircase}). On the plateaus the number of spin-$\uparrow$ atoms is insensitive to changes of the experimental parameters. Due to finite detection efficiency of the Rydberg atoms the plateaus do not show integer values. To confirm the number of excitations on a plateau one can look at the average spin-$\uparrow$ density which shows three spots for the stair of three spin-$\uparrow$ atoms (Fig.~\ref{fig:staircase} inset).
We note that although a macroscopic population of the many-body ground state is reached in these experiments the system is not prepared with high fidelity in the absolute ground state \cite{Petrosyan2016}. Reaching the absolute ground state of a mesoscopic quantum many-body system is challenging, and simulations show that obtaining near unit fidelity would be only possible if the sweep is longer than about \SI{100}{\micro\second} which is beyond the lifetime of the Rydberg state and therefore impossible in this experimental setting. Slight deviations from adiabaticity leak population to states very close in energy and the states populated in this way are experimentally effectively indistinguishable from the ground state. A typical example is the shift of one spin-$\uparrow$ by one site in a one-dimensional crystal.

\begin{figure}
\centering
\includegraphics{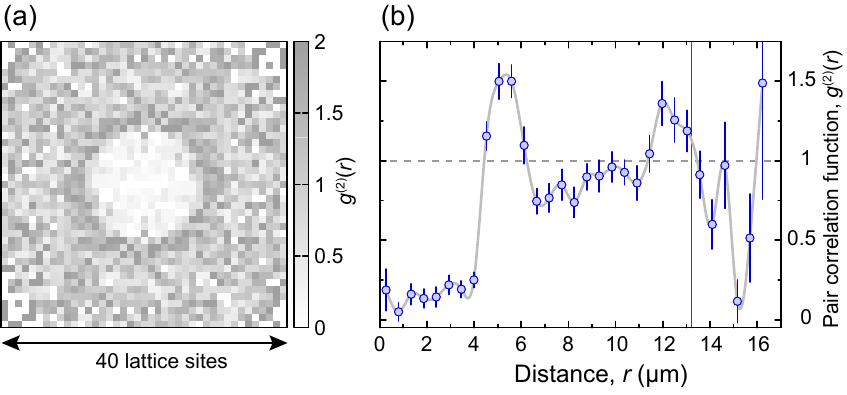}
\caption{Pair correlation in 2D after an adiabatic sweep. (a) Two-dimensional spatial correlation
function for a disc-shaped initial system with radius of 11.8(2) lattice sites which was excited by an adiabatic sweep (for the largest system size shown in ref.~\cite{Schauss2015}). There is a clear circular anti-correlation region in the center with a ring around. A second ring is barely visible due to the noise caused by limited statistics. Due to the symmetry in the correlation function the graph is point symmetric around the center. (b) Blue points show the radial correlation function calculated as defined in Eq.~\eqref{eq:g2r} for the data shown in (a). The red line marks the diameter of the system and the grey line is a guide to the eye.  \label{fig:spatial_correlation_munich_2d}} 
\end{figure}

\begin{figure}
\includegraphics{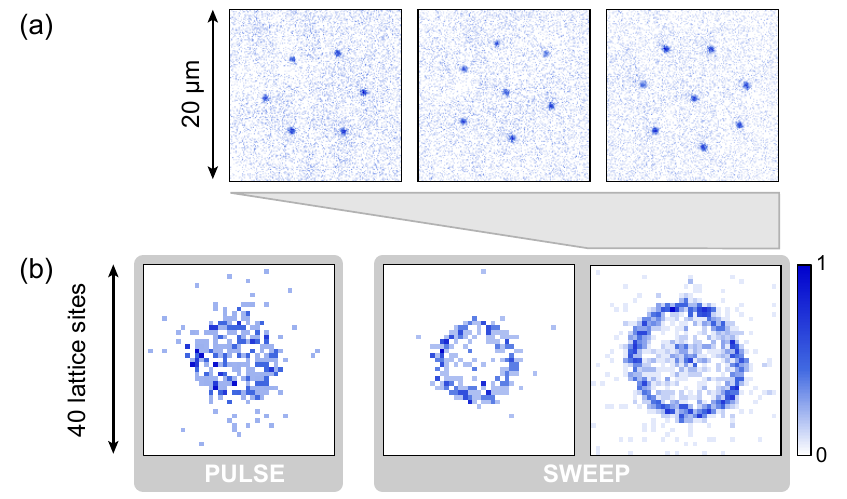}
\caption{Preparation of low-energy states in disc-shaped samples.
 (a) Unprocessed experimental single shot pictures with $6$, $7$
  and $8$ Rydberg atoms from the rightmost data set. Each blue point corresponds to a single atom. (b) Measured two-dimensional
  distribution of the magnetization. The colour scale represents the
  normalized counts per site. Magnetization densities for pulsed 
  (left grey box) and sweeped laser coupling with increasing system size from left to right are shown in the right grey
  box. The pulsed coupling was done with the same amplitude modulation as
  for the sweep, but the detuning $\Delta$ was held constant. Adapted from ref.~\cite{Schauss2015}. \label{fig:crystal_2d}}
\end{figure}

From these results in 1D directly the question arises if the deterministic preparation of ordered states can be also implemented in two-dimensional systems. Experimental parameters are much more challenging in 2D and it becomes hard to observe the staircase of excitations in the same lattice with 43S rubidium Rydberg states. This becomes obvious when looking at the interaction energy for the different configurations of different spin-$\uparrow$ numbers (Fig.~\ref{fig:crystal_energy}c). In the case of a circular initial system the radius needs to be adjusted to less than a lattice site to observe a plateau in the staircase. This is effectively impossible in an experimental setting with a preparation uncertainty of about 80\,\% filling in the initial state. The imperfect filling leads already to effective radius variations of the order of one site even if slight alignment errors of the addressing pattern with the lattice are neglected. But it is experimentally possible to prepare low energy configurations of the spin-$\uparrow$ numbers although the spin-$\uparrow$ number is not fully deterministic. One hint that lower energy states are reached is provided by comparing the correlation function (Fig.~\ref{fig:spatial_correlation_munich_2d}) from data with adiabatic sweep with the previously measured one without adiabatic sweep (Fig.~\ref{fig:spatial_correlation_munich}).
Another way is to directly look at the average magnetization density without any configuration alignment. Low energy states will show a ring-like structure in the spin-$\uparrow$ density which has not been observed for the simple pulsed excitation (Fig.~\ref{fig:crystal_2d}). When a state with mainly $n_\uparrow=3$ is excited the blockade radius requirement leads automatically to a hole in the spin-$\uparrow$ density in the center. This demonstrates that the adiabatic preparation is partly working in 2D but the precision and stability of the sweep and initial atom configurations are not sufficient to reach deterministic preparation of states with certain spin-$\uparrow$ number. The quality of the sweep in these experiments was mainly limited by fluctuations in the lightshift caused by laser intensity noise and decoherence due to a combined two-photon laser linewidth of approximately $\SI{50}{kHz}$. In addition the large spacing of the Rydberg crystals in these experiments are unfavorable for adiabatic preparation but experimental constraints did not allow to reduce the blockade radius significantly in the described setting without deteriorating detection efficiency.

\subsection{Challenges and Limits}

The vastly different energy scales of Rydberg atoms and ground state atoms in the optical lattice can lead to interesting physics but also pose experimental challenges. This starts with high-fidelity imaging of Rydberg atoms and also experiments as simple as determining the resonance frequency of the Rydberg transition can become complicated in the presence of strong interactions.
In a dense ensemble the Rydberg line exhibits significant shifts and broadening, requiring to work with very dilute atomic ensembles or Rydberg excitation numbers on the order of one to determine the Rydberg line with high precision.
But the position of the Rydberg line has to be determined very precisely for the adiabatic sweeps as the relative frequency offset of the sweep with respect to the resonance needs to be adjusted to few ten Kilohertz to excite certain excitation configurations. 
Another experimental issue is that Rydberg atoms experience a different potential landscape than ground state atoms in optical lattices. The conflicting requirements of targeted lattice spacing and necessary lattice wavelength for trapping of ground and Rydberg state make it very hard to design optical lattices with reasonable parameters that trap Rydberg and ground states of alkaline atoms in the same places. But for optical tweezers equal trapping for ground and Rydberg state (magic trapping) has been achieved \cite{Zhang2011,Maller2015}. For alkaline atoms in optical lattices the options are mainly to avoid the effects of light shifts on the Rydberg atom by minimizing the time spent in the Rydberg state, switching the potential off during Rydberg excitation, working with Rydberg states that experience less forces \cite{Topcu2013} or phase-shifting the lattice \cite{Anderson2011}. For alkaline-earth-like atoms magic trapping in optical lattices is promising \cite{Mukherjee2011}.
Many challenges for the preparation of crystalline configurations are caused by the very fast fall-off of the Rydberg van der Waals interaction with $1/r^6$, leading to stringent requirements for the precision of the size of the initial atom distribution and exact tuning of the frequency, as the interaction energy drops very fast to small values that still need to be resolved for the adiabatic preparation.
The very fast rise of the interaction energy for distances below the blockade radius leads to problems with energy scales as the energy pumped into the system by an adiabatic sweep can lead to strong forces between the Rydberg atoms that effectively heat up the system due to interactions between Rydberg atoms. In particular in optical lattices a slight heating or shift of the atoms leads already to occupation of higher bands in the lattice and therefore increased tunnelling during imaging of the Rydberg atom positions.
The adiabatic preparation as discussed here is fundamentally limited to small excitation numbers as the gaps in the adiabatic preparation drop exponentially with increasing atom number \cite{Pohl2010a} leading to immense requirements in terms of lifetime and coherence time in the experiments that cannot be fulfilled with the lifetime of typically used S/P/D Rydberg states.
The extension to anisotropic interactions for the creation of small crystals is possible \cite{Vermersch2015}. Current experimental techniques also allow for a detailed investigation of the other phases in the phase diagram which are theoretically more interesting than the mainly classical crystalline state, for example floating crystal phases \cite{Weimer2010b}. 
By optimizing time dependence of detuning and Rabi frequency by optimal control techniques it might become possible to not only create crystalline states with higher fidelity but also to prepare entangled states deterministically \cite{Cui2017}.

\section{Conclusion}

Recently there has been tremendous progress in controlling Rydberg excitation coherently and employing them to prepare entangled states of neutral atoms. The combination of Rydberg experiments in large ensembles with high-resolution imaging in optical lattices opens a wide range of new experiments in long-range interacting spin models. In this review we focussed on a finite-range Ising spin Hamiltonian that emerges naturally when exciting an ensemble of atoms on a lattice to a Rydberg state.

\subsection{Possible improvements}

One of the experimentally most important open questions about working with Rydberg atoms is a precise characterization of the sources for decoherence and dephasing. They are plentiful and it is hard to track down certain experimental uncertainties and eliminate them. This is partly caused by the sensitivity of Rydberg atoms to electromagnetic fields, which is inherently connected to their strong interactions. Static electric fields, microwave fields in the Gigahertz range and black body radiation have all a strong influence on Rydberg atoms and are rarely controlled in typical ultracold atom experiments due to their negligible influence on neutral atoms. Specialized Rydberg experiments control the electric fields but reduction of black body radiation is currently limited to only very few experiments. The effects that might be caused by black body radiation are still barely understood. Black body radiation effectively leads to diffusion of Rydberg state population to nearby Rydberg states. As many experiments cannot resolve population of neighbouring Rydberg states in detection, these repopulation effects show up as loss of Rydberg atoms in many experiments and effectively reduce detection efficiency.
Minimizing effects of black body radiation is challenging, the obvious solution is to cool the whole vacuum chamber to low temperatures. There might be ways to use Rydberg states with lower coupling to black body radiation. Selecting a special Rydberg state with particular neighbourhood of other states might lead to an improvement but the influence of the level splitting on interactions limits this approach. In some experiments, the use of additional lasers to depump unwanted Rydberg states that were populated by black-body radiation might be a solution.

There is also the general question to what extent the lifetime of systems with finite-range interactions can be improved. Typical interactions between Rydberg atoms lead to interaction forces on short scales larger than the trapping of the atoms which turn the system intrinsically unstable if Rydberg atoms can come close to each other. Additionally optical driving to these interacting states leads to extreme broadening of the excitation lines which becomes even worse when unwanted states are populated in the system \cite{Balewski2014,Goldschmidt2016,Aman2016,Zeiher2016}. This is in particular a problem in van-der-Waals-interacting systems which rely on second-order interactions where impurities in other Rydberg states can introduce first-order dipole-dipole interactions that are much stronger. Developing techniques to keep these systems under control still needs more detailed understanding. This will also increase understanding in other systems as some of the effects are quite generic for dipolar interacting systems. 

\section{Future prospects}

The investigation of spin models created by the coupling of atoms in lattices to Rydberg states just started and many fields have not been explored yet. 
But even many predicted effects in the Ising spin model discussed here have not been seen in experiments. In the following we discuss an exemplary set of open questions. 
\paragraph{Entanglement} One big open field is still the investigation of entanglement properties of mesoscopic Rydberg-excited states. This requires very good detection and preparation, and even then, typically a specifically designed entanglement witness is required to show entanglement in reasonably large systems.
Besides the demonstration of entanglement of two atoms \cite{Wilk2010,Isenhower2010,Jau2016} and the special case of the Rydberg superatoms \cite{Zeiher2015,Ebert2015} entanglement experiments with Rydberg atoms are scarce.
It has been proposed to generate spin squeezing and non-Gaussian states with Rydberg atoms which can be directly applied in optical lattice clocks \cite{Bouchoule2002,Opatrny2012,Gil2014,Macri2016}.
Lately also interest came up in the study of the entanglement growth in quench experiments \cite{Schachenmayer2013,Hauke2013}.
\paragraph{Phase diagram of the Ising model with finite-range interactions} The crystalline phase is only a small region of the phase diagram of the Ising model with finite-range interactions. A more detailed measurement of correlation functions and excitation fractions should allow to pin down phase transitions experimentally. This includes the dynamics of the build up of order and the investigation of the path to long-range order \cite{Hoening2014}.
\paragraph{Dipole interactions} Next to van-der-Waals interactions Rydberg atoms also allow for the investigations of spin models based on the direct dipole-dipole interaction between different Rydberg states. First experiments on the exchange interactions \cite{Barredo2015} look promising for the investigation of larger spin systems.
\paragraph{Exotic spin systems} Spin systems with exotic interactions can be designed with Rydberg atoms \cite{Glaetzle2015,VanBijnen2015}. These ideas widen the scope of Rydberg spin systems to a much larger class of spin models and employ the Rydberg dressing technique discussed in the following.

\subsubsection{Beyond spin models - Rydberg dressing}
One of the big targets is to implement interacting many-body systems combining atomic motion with tunable long-range interaction via Rydberg atoms \cite{Baranov2008}. This could be achieved by engineering a tunnelling term for the atoms in the lattice of the Ising model dicussed here. The main experimental challenge is to bridge the mismatch in energy and timescales between the Rydberg excitation and the dynamics of ground state atoms. A possible solution is the so-called Rydberg dressing where ground state atoms are coupled off-resonantly to Rydberg states leading to effectively weaker interaction with lower decay rates. 

Rydberg dressing is the off-resonant admixture of a Rydberg state to the ground state in the limit $\abs{\Delta} \gg \Omega$. But there is a continuous transition from off-resonant Rydberg excitation such that the description is similar also closer to resonance. The motivation for Rydberg dressing is to tune the lifetime of Rydberg atoms to another independent timescale like the tunnelling in the optical lattice. In this case the lifetime of the Rydberg-dressed ground state atom has to be long compared to the tunnelling in the lattice.  The main difficulty in this approach is that decay and loss processes of Rydberg atoms have to be controlled on these timescales that are much longer than for near-resonant experiments such that also more exotic loss processes become relevant.
Rydberg dressing has been proposed to implement interactions in quantum gases \cite{Johnson2010a,Honer2010,Pupillo2010,Glaetzle2012,Macri2014,Gaul2016,Sandor2016} together with techniques to detect weak dressing interactions \cite{Li2012,Gil2014,Mukherjee2016}. One of the exotic states that might be possible to realize via Rydberg dressing is a supersolid droplet crystal \cite{Pupillo2010,Henkel2010a,Cinti2010,Henkel2012,Macri2013,Cinti2014,Lechner2015}. Rydberg dressing also allows to implement local constraints that are at the heart of the implementation of models related to gauge theories like the quantum spin ice \cite{Glaetzle2014}. 
But in addition, it is also possible to design interaction terms in spin Hamiltonians that are quite unusual, for example terms that conserve the parity of the spin but not the magnitude \cite{Glaetzle2015,VanBijnen2015}. 
Another prediction are cluster Luttinger liquids in 1D \cite{Mattioli2013} and the implementation of glassy phases \cite{Olmos2012b,Lechner2013,Angelone2016}. It might be even possible to implement a universal quantum simulator \cite{Weimer2010a} and quantum annealer based on Rydberg dressing \cite{Lechner2015b,Glaetzle2016}.

Experimental implementation of Rydberg dressing turned out to be challenging and schemes based on two-photon excitation in large systems very close to resonance suffered from strong loss \cite{Balewski2014,Goldschmidt2016,Aman2016}.
Rydberg dressing has recently been implemented first with two atoms \cite{Jau2016} and in a many-body setting \cite{Zeiher2016}. Both of these experiments used a direct excitation to Rydberg states with an ultra-violet laser. This direct coupling scheme is superior in experimental parameters to two-photon excitation schemes in alkali atoms. An experimental implementation of many-body systems with finite-range interactions and hopping in the lattice would open another new field for Rydberg physics.

\section*{Acknowledgments}
\begin{acknowledgments}  
I thank I Bloch, S Kuhr and C Gross for support of the Rydberg project and supervision of my PhD thesis, J Zeiher, M Cheneau and A Omran for the joint work on the Rydberg experiments and the theory support by T Pohl and T Macr\`\i. I also thank C Weitenberg, M Endres, S Hild, T Fukuhara, J-y Choi, F See{\ss}elberg, D Bellem for their contributions to the experiments in Munich. I am grateful for comments on the manuscript by A Browaeys, C Gross, T Lahaye, A Omran, T Macr\`\i, and J Zeiher. P.S. was supported by a Dicke fellowship from Princeton University.
\end{acknowledgments}

\bibliography{bibliography,paper1,paper2,paper3}

\end{document}